\documentclass[prb,aps,twocolumn,showpacs,showkeys]{revtex4}

\usepackage{epsfig}
\usepackage{amsmath}
\usepackage{amsfonts}
\usepackage{bm}

\begin{document}

\title{Magnetic polarons in doped 1D antiferromagnetic
chain}

\author{I.~Gonz\'alez, J.~Castro, and D.~Baldomir}
\affiliation{Departamento de F\'{\i}sica Aplicada, Universidade de
Santiago de Compostela, E-15706 Santiago de Compostela, Spain}

\author{A.~O.~Sboychakov, A.~L.~Rakhmanov, and K.~I.~Kugel}
\affiliation{Institute for Theoretical and Applied
Electrodynamics, Russian Academy of Sciences, Izhorskaya Str.
13/19, Moscow, 125412 Russia}

\begin{abstract}

The structure of magnetic polarons (ferrons) is studied for an 1D
antiferromagnetic chain doped by non-magnetic donor impurities.
The conduction electrons are assumed to be bound by the
impurities. Such a chain can be described as a set of ferrons at
the antiferromagnetic background. We found that two types of
ferrons can exist in the system. The ground state of the chain
corresponds to the ferrons with the sizes of the order of the
localization length of the electron near the impurity. The ferrons
of the second type produce a more extended distortion of spins in
the chain. They are stable within a finite domain of the system
parameters and can be treated as excitations above the ground
state. The ferrons in the excited states can appear in pairs only.
The energy of the excited states decreases with the growth in
density of impurities. This can be interpreted as a manifestation
of an attractive interaction between ferrons.
\end{abstract}



\maketitle

\section{Introduction}

It is now commonly accepted that the tendency toward phase
separation is of fundamental importance for the physics of
manganites and related compounds~\cite{Nag,Dag,Kak}. The
self-trapping of charge carriers is the most widely discussed type
of phase separation, first predicted in the seminal paper of
Nagaev~\cite{Nag67}. In such phase-separated states, charge
carriers are confined within small ferromagnetic metallic droplets
(magnetic polarons or ferrons) located in an insulating
antiferromagnetic matrix. It is usually assumed that the region of
the perturbed spins around a ferron is rather narrow that is
confirmed by numerical calculations for some
models~\cite{PatSat,Klap}. However, as first pointed out by De
Gennes \cite{DeG60}, the distortion of the magnetic order around a
magnetic defect (e.g. a ferron) may decay slowly with the
distance. The possible existence of ferrons with such extended
spin distortions was analyzed for one-dimensional (1D)
antiferromagnetic chain in Refs.~\onlinecite{Nag01b,Art3}. These
calculations performed for an isolated impurity show that the
characteristic length of the distorted spin surrounding can be
much larger than the size of the trapping region.

The problem arises what kind of ferrons should be expected at
different values of the parameters of the system. To study this
problem, we consider in this paper 1D antiferromagnetic chain
doped with non-magnetic donor impurities of finite density. We
find that there are two possible magnetic structures: one
characterized by extended spin distortions and another one with a
narrow distorted region. For the relationship between parameters
characteristic of manganites, the ground state for a regular
distribution of impurities corresponds to the ferrons without
extended spin distortions. The ferrons with extended spin
distortions can appear as excited metastable states. We found that
due to the overlapping of extended spin distortions, an attractive
interaction among these ferrons arises in the system. This
interaction could favor a phase-separation process, that would
lead the system to an inhomogeneous state with clusters of
ferrons, as observed in the experiment.

\section{The model}

In this section, we find the magnetic structure of the ground
state of an 1D antiferromagnetic chain doped by non-magnetic donor
impurities. We consider an one-dimensional chain of
antiferromagnetically coupled local spins, the $y$-axis being the
axis of the chain. We do not considered here specific effects
related to one-dimensionality and start from the antiferromagnetic
structure characteristic of the mean-field approximation.
Non-magnetic impurities occupy the sites of the chain at
half-integer positions. The conduction electrons are assumed to be
in a bound state of the impurity electrostatic potential. It is
also assumed, for simplicity, that the wave function of the
conduction electron extends only over the two neighbouring
magnetic ions near an impurity. Further on, we assume that for a
given density of conduction electrons $n$, the impurities are
periodically distributed along the chain, and $L=1/n>2$ being the
distance between neighboring impurities (in units of the lattice
constant). We consider the following Hamiltonian of the system

\begin{equation}\label{eq:1a}
H = H_{\text{el}}+J'\sum\limits_{g}\vec{S_{g}}\vec{S}_{g+1} -K'
\sum\limits_{g}\left(S^{x}_{g}\right)^{2}\ ,
\end{equation}
where
\begin{eqnarray}\label{eq:1b}
H_{\text{el}} &=& -t\sum\limits_{i,\sigma}
\left(a^{+}_{iL,\sigma}a_{iL+1,\sigma}+a^{+}_{iL+1,\sigma}a_{iL,\sigma}\right)-
\nonumber\\ & & -\frac{A}{2} \sum\limits_{i,\sigma,\sigma'}\left\{
a^{+}_{iL,\sigma}\left(\vec{\sigma}
\vec{S}_{iL}\right)_{\sigma,\sigma'}
a_{iL,\sigma'}+\right.\nonumber\\ & &\left.
+a^{+}_{iL+1,\sigma}\left(\vec{\sigma}
\vec{S}_{iL+1}\right)_{\sigma,\sigma'} a_{iL+1,\sigma'} \right\}
\end{eqnarray}
In Eqs. (\ref{eq:1a}) and (\ref{eq:1b}), $\vec{S_{g}}$ is the spin
of the magnetic ion located at site $g$, treated as classical
vector, symbols $a^{+}_{g,\sigma}$, $a_{g,\sigma}$ denote the
creation and annihilation operators for the conduction electron
with spin $\sigma$ at site $g$, and $\vec{\sigma}$ are Pauli
matrices. The second and third terms in Eq. (\ref{eq:1a}) are the
antiferromagnetic exchange between local spins and the magnetic
anisotropy energy, respectively. The two terms in $H_{\text{el}}$
describe the kinetic energy of the conduction electrons bounded by
the impurities, which are located between sites with $g=iL$ and
$g=iL+1$ ($i$ is an integer), and the Hund-rule coupling between
the conduction electrons and the localized spins. Parameters $A$,
$t$, $J'$, and $K'$ of Hamiltonian~(\ref{eq:1a}) are considered to
be positive and the $x$ axis is the easy axis. The energy of
Coulomb interaction between the conduction electron and the
impurity is an additive constant and, for this reason, omitted in
the calculation.

Hamiltonian~(\ref{eq:1a}) is applicable for the description both
of wide-band and double-exchange (like manganites) magnetic
semiconductors. The hierarchy of parameters in wide-band case is
$t\gg AS>J'$ while for the double-exchange case we have $AS\gg
t>J'$. In both cases $K'$ is assumed to be the smallest parameter,
which corresponds in the most cases the experimental situation.

It is well known that due to the Hund-rule coupling $A$
Hamiltonian (\ref{eq:1b}) favors the deviations from the ideal
antiferromagnetic arrangement of spins. The deviations are the
most pronounced for the sites neighboring to impurities. This can
be illustrated by the diagonalization of Hamiltonian
$H_{\text{el}}$ (\ref{eq:1b}), which can be easily performed. As a
result, we have that the lowest eigenvalue of $H_{\text{el}}$
corresponding to the ground state and the low-lying excitations
can be written as

\begin{equation}\label{eq:2}
E_{\text{el}}=-\frac{1}{2}\sum\limits_{ i}
\sqrt{A^{2}S^{2}+4t^{2}+4ASt\cos\left(\frac{\nu_i}{2}\right)}\ ,
\end{equation}
where $\nu_i$ is the angle between vectors $\vec{S}_{iL}$ and
$\vec{S}_{iL+1}$. We can see that the lowest value of energy
$E_{\text{el}}$ corresponds to the parallel orientation of spins
nearest to the impurity. So, we get a ``seed" ferron centered at
the impurity.

To the first approximation, both wide-band and double-exchange
cases can be treated in the same manner. At $t\ll AS$, the
first-order term in $E_{\text{el}}$ with respect to $t/AS$ has the
form

\begin{equation}\label{eq:2b}
E_{\text{el}}\approx-\frac{AS}{2}N_{\text{imp}} -t\sum\limits_{
i}\cos\left(\frac{\nu_i}{2}\right)\ ,
\end{equation}
where $N_{\text{imp}}$ is the number of impurities. In the case
$t\gg AS$, the similar expression can be obtained by interchanging
$t\longleftrightarrow AS/2$. Further on, we discuss the double
exchange case (\ref{eq:2b}), however, the results for the
wide-band case can be obtained by the aforementioned change of the
parameters.

Let us consider now the configuration of the local spins
corresponding to the ground state. Due to the rotational symmetry
of the Hamiltonian (\ref{eq:1a}) with respect to the easy axis,
all local spins in the ground state should lie in a plane
containing the easy axis. The deviations of any local spin from
this plane results in the growth of the energy of the system. We
assume without loss of generality that it is the $xz$ plane.

\begin{figure}
\begin{center}
\epsfig{file=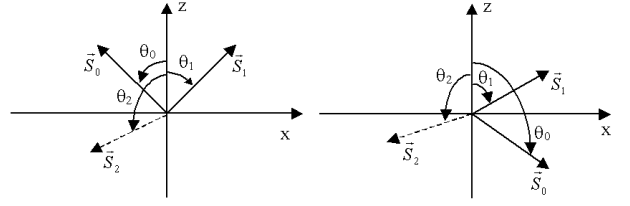,width=8cm}
\end{center}
\caption{\label{fig:1} The directions of  local spins at sites
with $g=0, 1$, and $g=2$, corresponding to the first (left panel)
and second (right panel) configuration of  local spins. The
symmetry condition for the first (second) case is
$\theta_1=\theta_0$ ($\theta_1=\pi-\theta_0$).}
\end{figure}

The direction of a local spin $\vec{S}_{g}$ in the $xz$ plane can
be characterized by angle $\theta_g$ with respect to $z$ axis (see
Fig.~\ref{fig:1}). Then, the spins located between any two
neighboring impurities can be represented as $\vec{S}_{g}= S
\left(\left(-1\right)^{g-s}
\sin\theta_{g},0,\cos\theta_{g}\right)$, where
$0\leq\theta_{g}\leq\pi$, and $s=0$ or 1. The passing through a
ferron changes such a sequence of spins to another one. Since the
impurities are periodically arranged along the chain, the magnetic
ordering of local spins in the ground state has to be also
periodic. There exist several ways to join the solutions at
impurities, which preserve the periodicity. The corresponding
relationships can be written as $S_{iL}^{xz} = \pm S_{iL+1}^{xz}$.
However, only two of them give the energy gain and correspond to
the ferron-type solutions. The first one is
$S_{iL}^{x}=-S_{iL+1}^{x}$, $S_{iL}^{z}=S_{iL+1}^{z}$. This means
that the magnetic moments of ferrons are directed along the
$z$-axis. In the second case, the magnetic moments of ferrons are
directed along the easy axis, and spins close to the $i$th
impurity satisfy the equality $S_{iL}^{x}=S_{iL+1}^{x}$,
$S_{iL}^{z}=-S_{iL+1}^{z}$. The configuration of local spins
corresponding to the first case is
$\vec{S}_{g}=S\left(\left(-1\right)^{g-1}\sin\theta_{g},0,\cos\theta_{g}\right)$
for all $g$, and the symmetry condition is
$\theta_{iL}=\theta_{iL+1}$. The configuration of  local spins
corresponding to the second case can be written in the form
$\vec{S}_{g}=
S\left(\left(-1\right)^{g-1-i}\sin\theta_{g},0,\cos\theta_{g}\right)$
for $iL+1\leq g\leq (i+1)L$, and the symmetry condition in this
case is $\theta_{iL}=\pi-\theta_{iL+1}$. Both configurations are
schematically shown in Fig.~\ref{fig:1}.

In the 1D antiferromagnetic chain under study the magnetic moments
of neighboring ferrons are either parallel or antiparallel for odd
and even $L$, respectively. If $L$ is odd integer, the periodicity
condition is $\theta_{g+L}=\theta_{g}$, and the magnetic moments
of all ferrons are parallel each other. For even $L$, the
periodicity condition is $\theta_{g+L}=\pi-\theta_{g}$ and the
magnetic moments of neighboring ferrons are antiparallel.

Taking into account the periodicity of the problem and the
aforementioned symmetry conditions, we readily come to the
relationships between the canting angles $\nu_i$ in Eqs.
(\ref{eq:2}) and (\ref{eq:2b}) and the angles $\theta_{iL}$,
$\theta_{iL+1}$. For any $L$, these relationships can be written
in the following form:
\begin{equation}\label{eq:nu}
\cos\left(\frac{\nu_i}{2}\right)=\left|\cos\left(\frac{\theta_{iL}
\pm\theta_{iL+1} }{2}\right)\right|\ ,
\end{equation}
where plus (minus) sign corresponds to the first (second)
configuration.

Now let us find the form of magnetic ordering for local spins and
the energy of the system corresponding to each configuration.
Substituting the expressions for $\vec{S}_{g}$ and Eq.
(\ref{eq:nu}) for canting angles $\nu_{i}$, into the Eqs.
(\ref{eq:1a}) and (\ref{eq:2b}), we obtain the energy of the
system in the following form

\begin{eqnarray} \label{eq:3}
E=J \sum\limits_{ g\neq iL}\cos
\left(\theta_{g}+\theta_{g+1}\right) -K \sum\limits_{ g}
\sin^{2}\theta_{g}+ \nonumber\\ \sum\limits_{ i} \left(J\cos
\left(\theta_{iL}\pm \theta_{iL+1}\right)-
2F\left|\cos\left(\frac{\theta_{iL}\pm
\theta_{iL+1}}{2}\right)\right|\right)\ ,
\end{eqnarray}
where we use the notation $J=J'S^{2}$, $K=K'S^{2}$, and $F=t/2$ in
the case $t\ll AS$, or $F=AS/4$ for $t\gg AS$. The plus (minus)
sign in Eq.~(\ref{eq:3}) corresponds to the first (second)
configuration of local spins. The constant term in (\ref{eq:2b})
is omitted here.

To find the spin structure of the ground state, one has to
minimize the energy (\ref{eq:3}) with respect to the angles
$\theta_{g}$. A set of nonlinear equations is obtained

\begin{widetext}
\begin{eqnarray} \label{eq:4}
\begin{cases}
J\sin\left(\theta_{g}+\theta_{g+1}\right)+
J\sin\left(\theta_{g-1}+\theta_{g}\right)
+K\sin\left(2\theta_{g}\right)=0 \qquad \qquad \qquad \qquad
\qquad \qquad g\neq iL+1,\left(i+1\right)L & \\
J\sin\left(\theta_{iL+1}+\theta_{iL+2}\right)\pm
J\sin\left(\theta_{iL}\pm \theta_{iL+1}\right)
+K\sin\left(2\theta_{iL+1}\right)=
2F\frac{\displaystyle{\partial}}{\displaystyle{\partial\theta_{iL+1}}}
\cos\left(\frac{\displaystyle{\nu_i}}{\displaystyle{2}}\right) &
\\
J\sin\left(\theta_{\left(i+1\right)L}\pm
\theta_{\left(i+1\right)L+1}\right)+
J\sin\left(\theta_{\left(i+1\right)L-1}+\theta_{\left(i+1\right)L}\right)
+K\sin\left(2\theta_{\left(i+1\right)L}\right)=
2F\frac{\displaystyle{\partial}}
{\displaystyle{\partial\theta_{\left(i+1\right)L}}}
\cos\left(\frac{\displaystyle{\nu_{\left(i+1\right)}}}
{\displaystyle{2}}\right) &
\end{cases}
\end{eqnarray}
\end{widetext}

Using the periodicity conditions
$\theta_{iL}=\theta_{\left(i+1\right)L}$,
$\theta_{\left(i+1\right)L+1}=\theta_{iL+1}$ for odd $L$, or
$\theta_{iL}=\pi -\theta_{\left(i+1\right)L}$,
$\theta_{\left(i+1\right)L+1}=\pi-\theta_{iL+1}$ for even $L$, the
number of equations in system (\ref{eq:4}) becomes equal to $L$.
After this, the system of equations is solved numerically at
different values of $F/J$ and $K/J$ assuming that $K/J \ll F/J,1$.
As a result, we obtain the magnetic structure of the system of
local spins with donor impurities. This structure is discussed in
the next section.

\section{Results}

\begin{figure}
\begin{center}
\epsfig{file=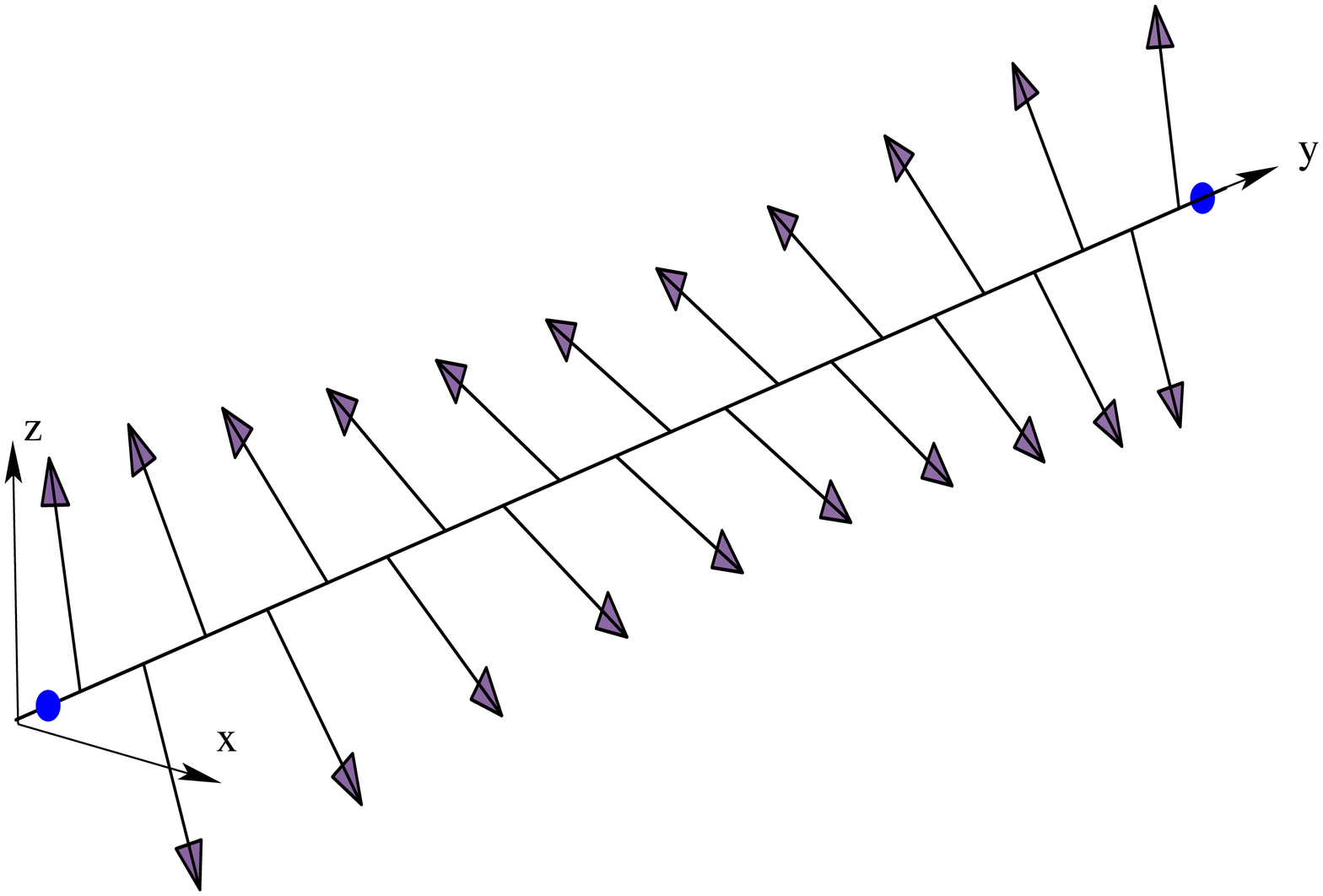,width=8cm}
\end{center}
\caption{\label{fig:2} Magnetic ordering for the first
configuration of local spins between two non-magnetic impurities
(dots) for $n=1/21$. This structure is repeated between any two
non-magnetic impurities of the chain. It can be seen that in the
region occupied by the conduction electrons (the first and the
last magnetic sites), we have almost ferromagnetic ordering, while
in the remaining part of the chain, there exists a distorted
antiferromagnetic ordering. The parameters of the Hamiltonian are
$F=3$, $K=2.5\,10^{-2}$ (both in $J$ units).}
\end{figure}

The numerical analysis of the solutions of Eqs.~(\ref{eq:4})
demonstrates that two types of ferrons schematically shown in
Fig.~\ref{fig:1} can exist in the system under study. For
relatively large $F/J\gtrsim1.5$ the ground state corresponds to
the ferrons of the second type (see right panel in
Fig.~\ref{fig:1}). For $F/J\gtrsim2$, we have a classical ferron
configuration~\cite{Nag67}, namely, ideal ferromagnetic ordering
inside the localization region of the doped electron (two lattice
constants in our case) and ideal antiferromagnetism outside this
region. The solution corresponding to the first configuration of
the local spins (see left panel in Fig.~\ref{fig:1}) is
characterized by a long-range deviation from the ideal
antiferromagnetic ordering and can be considered as an excitation
above the ground state. This spin structure is illustrated in
Fig.~\ref{fig:2}, where we plot the directions of local spins
between two impurities for $n=1/21$.

For smaller $F/J$, the ground state of the system corresponds to
the first type of ferrons. However, in this limit the difference
between the two types of ferrons is not so drastic. Both types of
ferrons create a distortion of antiferromagnetic ordering outside
the electron localization region. Moreover, the characteristic
size of the distorted region is larger for the ferrons of the
second type. Therefore, we come to the conclusion that for any
value of the ratio $F/J$ the ground state of the system
corresponds to the ferrons with smaller radius of spin
distortions.

\begin{figure}
\begin{center}
\epsfig{file=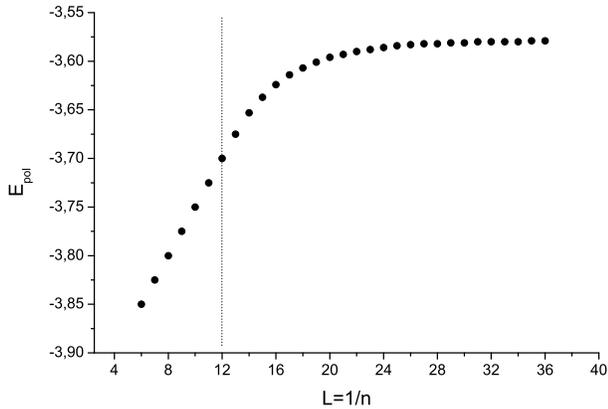,width=8cm}
\end{center}
\caption{\label{fig:3} Energy of ``coated" ferron versus the
inverse density of impurities $L=1/n$. The values of parameters
are  $F=3$, and $K=2.5\,10^{-2}$ (in $J$ units). The energy of a
``bare" ferron equals to $E_{\text{pol}}=-2(F-J)=-4$. At $L\leq
L_{\text{cr}}=1/n_{\text{cr}}=12$ (vertical line), the ``coated"
ferrons become unstable (see figures and text below).}
\end{figure}

\begin{figure}
\begin{center}
\epsfig{file=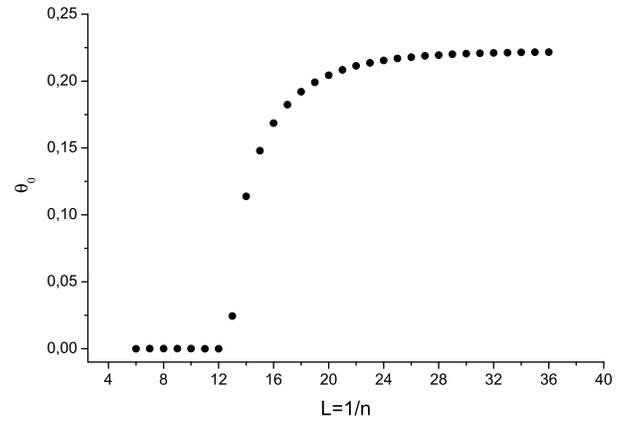,width=8cm}
\end{center}
\caption{\label{fig:4} Angle $\theta_{0}$ of the local spin and
the $z$ axis in the trapping region  versus the inverse density of
impurities, $1/n$. This angle is one-half of the canting angle
between spins in the trapping region. When the density of
impurities is greater than or equal to the critical value
$n_{\text{cr}}=1/12$, the canting angle is close to zero, and the
state corresponding to ``coated" ferrons becomes unstable. This is
a result of the overlapping of the spin distortions caused by
neighboring ferrons. The values of the parameters in the
Hamiltonian are $F=3$, $K=2.5\,10^{-2}$ (in $J$ units).}
\end{figure}

Let us consider in more detail the case $F/J>2$ (typical of
manganites). Let us refer the ferrons with and without extended
spin distortions as ``coated" and ``bare", respectively. The
system of ``bare" ferrons forms the ground state, whereas, the
system of ``coated" ferrons can form a metastable state. In
Figs.~\ref{fig:3} and \ref{fig:4}, we show the energy of the
``coated" ferron and the angle of the local spin in the trapping
region, $\theta_{0}$ as functions of the inverse impurity density,
$1/n$. The ferron energy $E_{\text{pol}}$ is defined as the
difference between energies of the system with and without doped
electrons divided by the number of impurities, that is,
$E_{\text{pol}}=E/N_{\text{imp}}+L(J+K)$. The energy of the
``bare" ferron does not depend on the density of impurities (up to
$1/n=1/2$) and equals to $E_{\text{pol}}=-2(F-J)$.

\begin{figure}
\begin{center}
\epsfig{file=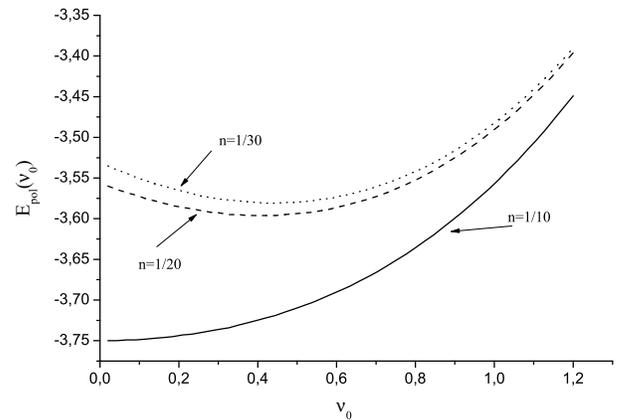,width=8cm}
\end{center}
\caption{\label{fig:5} Energy of the ``coated" ferron versus
canting angle at different impurity densities. When $n\geq
n_{\text{cr}}$ the $E_{\text{pol}}$ is a monotonic function of
$\nu_0$, and it has a minimum at the angle $\nu_0=0$ corresponding
also to the system of ``bare" ferrons, which has a lower energy.
This means that the system of ``coated" ferrons is unstable,
because there is no energy barrier between these two states.}
\end{figure}

In Fig.~\ref{fig:5}, we plot the energy of the ``coated" ferron as
a function of the canting angle $\nu_0$ at different impurity
densities. At small densities the minimum of the ferron energy
corresponds to nonzero value of $\nu_0$, whereas at the densities
exceeding some critical value $n_{\text{cr}}$ the ferron energy
has a minimum at $\nu_0=0$.

The numerical analysis of Eqs. (\ref{eq:4}) shows that for
impurity densities higher or equal than some critical density
$n_{\text{cr}}$, the angle $\theta_0$ (which is one-half of the
canting angle $\nu_0$) for the ``coated" ferrons  is equal to zero
(see Fig.~\ref{fig:4}). Moreover, all spins in the chain become
perpendicular to the easy axis. Thus, at $n\geq n_{\text{cr}}$,
the periodic structure of ``coated" ferrons becomes unstable.
Indeed, such ferrons become structurally identical to ``bare"
ferrons but have higher energy. Any rotation of this spin
structure toward the easy axis leads to the monotonic decrease of
its energy. In other words, at small densities each ``coated"
ferron creates a long-range spin distortion of the
antiferromagnetic background around it. This long-range spin
distortion stabilizes the ``coated" ferrons. The radius of
extended spin distortion created by a single ``coated" ferron can
be estimated as $r_{0}\approx 1/2n_{\text{cr}}$. When the density
increases, the distortions created by neighboring ferrons start to
overlap, the corresponding energy barrier lowers, and disappears
at $n=n_{\text{cr}}$.

The critical density increases with $K$ since any deviation of the
spins from the easy axis become less favorable with the growth of
anisotropy energy. The value of $n=n_{\text{cr}}$ decreases with
the increase of ratio $F/J$ since the smaller canting angles
become more favorable with the growth of the electron kinetic
energy $t$ compared with the exchange integral $J'$. Note that for
small values of $F/J\lesssim1$ both types of ferrons remain stable
up to the density $n=1/2$.

Let us consider now the low density limit $n\to 0$. As can be seen
from Figs.~\ref{fig:3} and \ref{fig:4}, the ferron energy and
angle $\theta_0$ are practically density independent at $n\ll
n_{\text{cr}}$. Therefore, in zero approximation, we can consider
``coated" ferrons as isolated objects. The numerical analysis
shows that the difference between energies of ``coated" and
``bare" ferrons $\Delta E$, weakly depends on the ratio $F/J$.
Following Ref.~\onlinecite{Art3}, it can be shown in continuous
approximation that $\Delta E\simeq\sqrt{8KJ}$, which is confirmed
by computational results. Since the ground state of the system
corresponds to ``bare" ferrons, the ferrons with extended spin
distortions can be considered as elementary excitations with the
energy $\Delta E$. Note, that these excitations can appear only in
pairs since a creation of a single ``coated" ferron strongly
disturbs the ground state and costs the energy at least of the
order of $J$. It is easy to see that to create a pair of
neighboring ``coated" ferrons, one needs to overcome the energy
barrier of the order of $K/n$ (because it is necessary to change
the orientation of all spins between two ferrons with respect to
easy axis). So, at low densities, this pair can be considered as a
metastable state.

The energy of a ferron in excited state $\Delta E$ is proportional
to the square root of $K$ and, for the case $K\ll J$, it can be
rather small in comparison with $J$. Therefore, even at low
temperatures $T\ll T_N\sim J$, one can expect to have a relatively
large number of ferrons in excited state. Using conventional
formulas for thermal averages and having in mind that ``coated"
ferrons arise in pairs, we have for the averaged density $n_1$ of
excited ferrons $\langle n_1\rangle_{T}=n/(1+\exp\{\Delta
E/kT\})$. This approach is valid only when the density of
impurities is less than $n_{\text{cr}}$, otherwise the ferrons
could not be considered as weakly interacting particles. However,
within the order of magnitude the estimates for $\langle
n_1\rangle_{T}$ and $\Delta E$ are valid up to $n\simeq
2n_{\text{cr}}$. Then, at $n>2n_{\text{cr}}$ the excitations under
discussion become unstable if the temperature exceeds a certain
critical value $kT_{\text{cr}}\simeq\Delta
E/\ln(n/n_{\text{cr}}-1)$. At $T>T_{\text{cr}}$ the density of
``coated" ferrons exceeds $n_{\text{cr}}$.

The ``coated" ferrons can be treated as isolated objects only in
the limit of $n\to 0$. At finite $n$, it is necessary to take into
account the density dependence of the ferron energy. Since the
energy of the ``coated" ferrons decreases with the growth of their
density, the interaction between them is attractive. This
interaction arises due to the overlapping of the spin distortions
of neighboring ferrons and its characteristic length far exceeds
the lattice constant. Although the calculation presented in
Fig.~\ref{fig:3} is valid only for a periodic distribution of the
impurities, it is natural to expect that the similar attractive
interaction should take place for random distribution of ``coated"
ferrons. The detailed calculation of an attractive interaction
between ferrons in a general case is the subject of further
research. We just note here that this attractive interaction can
be, in principle, the possible mechanism for formation of a
long-range phase-separated state involving large clusters of
ferrons.

\section{Conclusion}

We studied the magnetic structure for 1D antiferromagnetic chain
doped by homogeneously distributed non-magnetic donor impurities.
The ground state of the system corresponds to a set of ferrons
with characteristic size of the order of the localization length
of electrons near the impurity. The elementary excitations in the
system are ferrons with extended long-range spin distortions.
These excitations can be unstable for some range of parameters of
the system, in particular, at sufficiently high impurity density.
An attractive interaction exists between ferrons in excited state
due to the overlapping of the extended spin distortions of
neighboring ferrons. This attractive interaction can lead to a
phase-separated state with large clusters of ferrons.

\section*{Acknowledgments}

We are grateful to M.~Yu.~Kagan, D.~I.~Khomskii, A.~ V.~Klaptsov,
and F.~V.~Kusmartsev  for helpful discussions.

This work was supported by the Russian Foundation for Basic
Research (projects 02--02--16708, 03--02--06320, and
NSh-1694.2003.2), INTAS (project 01--2008), CRDF (project
RP2--2355--MO--02).

\end{document}